\documentclass[amsmath,showpacs,twocolumn,aps,prl,superscriptaddress]{revtex4}

\usepackage{amsmath,amsfonts,amssymb}
\usepackage{graphicx}
\usepackage{color}
\usepackage{enumerate}
\usepackage{bbm}

\makeatletter

\newcommand{\Rmnum}[1]{\expandafter\@slowromancap\romannumeral #1@}
\makeatother

\newcommand{\be}{\begin{equation}}
\newcommand{\ee}{\end{equation}}
\newcommand{\bea}{\begin{eqnarray}}
\newcommand{\eea}{\end{eqnarray}}

\newcommand{\I}{\mathbbm{1}}
	
\date{\today}

\begin{document}

\title{Structure factor of interacting one-dimensional helical systems}

\author{Suhas Gangadharaiah}
\affiliation{Department of Physics, Indian Institute of Science Education and Research, Bhopal, India}

\author{Thomas~L.~Schmidt}
\affiliation{Department of Physics, University of Basel, Klingelbergstrasse 82, 4056 Basel, Switzerland}

\author{Daniel Loss}
\affiliation{Department of Physics, University of Basel, Klingelbergstrasse 82, 4056 Basel, Switzerland}


\begin{abstract}
We calculate the dynamical structure factor $S(q,\omega)$ of a weakly interacting helical edge state in the presence of a magnetic field $B$. The latter opens a gap of width $2B$ in the single-particle spectrum, which becomes strongly nonlinear near the Dirac point. For chemical potentials $|\mu| > B$, the system then behaves as a nonlinear helical Luttinger liquid, and a mobile-impurity analysis reveals interaction-dependent power-law singularities in $S(q,\omega)$. For $|\mu| < B$, the low-energy excitations are gapped, and we determine $S(q,\omega)$ by using an analogy to exciton physics. \end{abstract}


\maketitle

The edge states of two-dimensional topological insulators (quantum spin Hall insulators) are gapless one-dimensional (1D) eigenmodes of the helicity operator, in which the spin orientation of a particle is correlated with its momentum~\cite{kane,bernevig,hasan,qi}. Electron-electron interactions within the edge modes are typically described using helical Luttinger liquid (HLL) theory~\cite{wu,xu,tanaka,das,braunecker}. Experimental evidence for helical modes was found via the observation of a quantized conductance $2e^2/h$ in HgTe/CdTe quantum wells~\cite{konig} and InAs/GaSb heterostructures \cite{knez11,suzuki13}. A crucial aspect of these edge states is that they host an odd number of Kramers doublets.
Consequently, unless time-reversal symmetry (TRS) is broken, the
gapless edge states remain robust against impurity scattering for not too strong interactions~\cite{xu,wu}.
Recently, the effect of  TRS breaking on transport~\cite{maciejko,maestro,schmidt13}, for the spin susceptibility~\cite{zak,klinovaja,stano,meng} and Coulomb drag between two HLL~\cite{zyuzin}, and the effect of interactions on backscattering via inelastic scattering channels have been explored~\cite{schmidt,budich}.

In this work we study the interplay of interactions and broken TRS on the dynamical structure factor (DSF) $S(q,\omega)$ (i.e., the Fourier transform of the density-density correlation function) of 1D helical fermions at zero temperature. In the absence of magnetic fields, the single-particle spectrum is linear, so $S(q,\omega)$ exhibits delta-function singularities at the mass shell $\omega = u|q|$, where the sound velocity $u$ depends on the interaction strength. In contrast, the magnetic field opens a gap and thus introduces a nonlinearity in the single-particle spectrum. This leads to changes in $S(q,\omega)$ which are not captured by the conventional HLL approach. These modifications are particularly nontrivial in the presence of interactions and manifest themselves as power-law singularities whose origin can be traced back to the Fermi edge singularity problem~\cite{schotte,balents,pustilnik,perira,zvonarev,mishchenko,karimi,matthews,imambekov09,imambekov}.

We begin our study by first discussing the DSF for the noninteracting case. The Hamiltonian density in the presence of a magnetic field $\vec{B}=(-B,0,0)$ perpendicular to the spin quantization axis of helical edge modes reads
\begin{align} \label{H1}
\mathcal{H}=-i \hbar v_F  \partial_x \sigma_3  + B \sigma_1-\mu,
\end{align}
where $v_F$ (henceforth we set  $v_F=1$ and $\hbar=1$) is the Fermi velocity, $\sigma_{1 ,3}$ are Pauli matrices, and $\mu>0$ is the chemical potential. The magnetic field $B$ (henceforth we will assume $B>0$) gaps out the single particle spectrum given by  $\epsilon_{\pm}(p)  = \pm \sqrt{p^2 + B^2}$, where $p$ is the momentum. The DSF $S(q,\omega)$ is a measure of the rate of formation of particle-hole pairs due to the absorption of an external excitation with momentum $q$ and frequency $\omega$. It is convenient to calculate the imaginary part of the retarded polarization operator, $\text{Im}\Pi^R(q,\omega)$, which is related to $S(q,\omega)$ via the fluctuation-dissipation theorem, $S(q,\omega>0) = - 2 \text{Im}\Pi^R(q,\omega)$. The energy-momentum representation of the zero-temperature polarization
operator in the Matsubara formalism is given by
\begin{align} \label{PO1}
\Pi^M(q,\omega)=\int\frac{d\epsilon dp}{(2\pi)^2}\mathrm{Tr} \Big[\mathcal{G}(p,i\epsilon)\mathcal{G}(p+q,i\epsilon+i\omega)\Big],
\end{align}
where $\mathrm{Tr} $ denotes the trace, the single-particle Matsubara Green's function is $\mathcal{G}(p,i\epsilon) =(1/2)\sum_{\beta=\pm}
(\I+ \beta \sigma_r)/(i\epsilon-\beta |\vec{r}| +\mu)$, and $\sigma_r = \vec{r}\cdot \vec{\sigma}/|\vec{r}|$ is the projection of the effective magnetic field $\vec{r} = B \hat{e}_x + p\hat{e}_z$ on the vector of Pauli matrices~\cite{gangadharaiah}. Performing the frequency integration and analytic continuation to real frequencies yields,
\begin{align} \label{PO2}
 & S(q,\omega)=\frac{1}{2}\int_{-\infty}^{\infty}dp\sum_{\beta,\beta'=\pm}\left(1+\beta\beta' \frac{\vec{r}\cdot\vec{s}}{|\vec{r}||\vec{s}|}
 \right)\\
 &\times \Big[\Theta(\mu-\beta |r|) -\Theta(\mu-\beta' |s|) \Big]\delta(\omega +\beta |\vec{r}| -\beta' |\vec{s}|), \notag
\end{align}
where $\vec{s}=  B \hat{e}_x + (p+q) \hat{e}_z$ and $\Theta(x)$ is the Heaviside function.

Let us first discuss the case $\mu > B$. In that case, the absorption spectrum ($\omega>0$) receives contributions from the following values of $(\beta,\beta')$: $(+,+)$ which corresponds to (intra-band) transitions within the upper band, and  $(-,+)$ which signifies (inter-band) transitions from the lower to the upper band.

For $B = 0$ the electronic spectrum is linear and spin is a good quantum number. Therefore, only spin-conserving transitions are allowed and Eq.~(\ref{PO2}) becomes a delta-function, $S(q,\omega)=[|q|    +  (|q|-\mu ) \Theta(|q|-\mu )]\delta(\omega - |q|)$.

\begin{figure}[t]
            \includegraphics[width=\columnwidth]{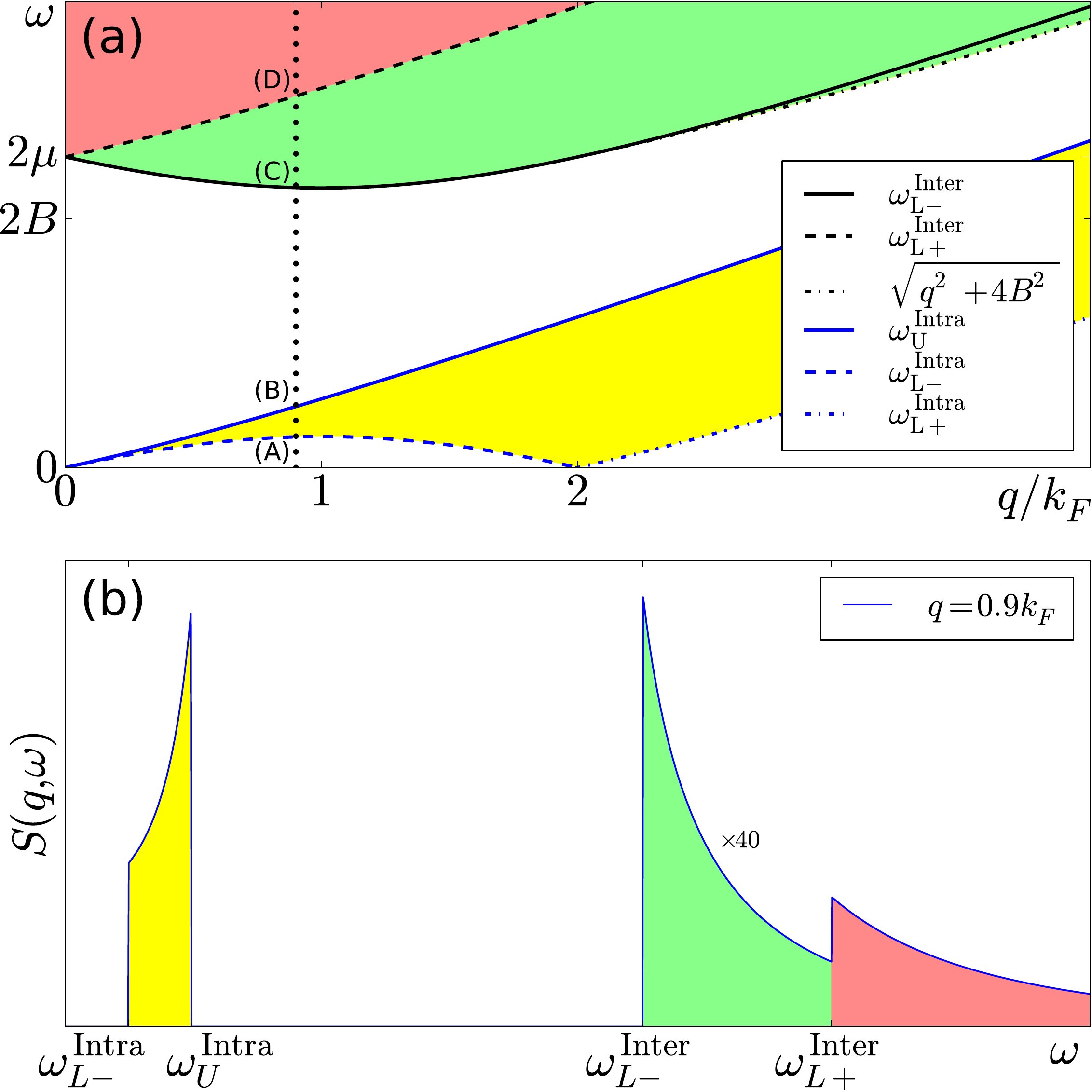}
    \centering
    \caption{(a) Regions in the $(q,\omega)$ plane where the noninteracting dynamical structure factor is nonzero.
      Yellow (red and green) regions denote contributions from
    the intra-band (inter-band) transitions. The vertical dotted line at $q=0.9 k_F$ indicates the position of the cut of $S(q,\omega)$ for fixed momentum.
     (b) The dynamical structure factor as function of $\omega$ (for fixed $q=0.9 k_F$) exhibits
    discontinuities at the edge thresholds.  }
    \label{fig:DSF}
\end{figure}

On the other hand, a nonzero magnetic field $B$ relaxes the constraints on the transitions and additional regions of support emerge in the $(\omega,q)$ plane. These non-overlapping regions can be classified as those acquiring contributions from either intra-band or  inter-band transitions, see  Figs.~\ref{fig:DSF}a and~\ref{fig:DSF}b. The total DSF is a sum of contributions from these two region, $S = S_1 + S_2$.

\emph{Intra-band transitions}: The contribution to the DSF from
the intra-band transitions is
\begin{eqnarray}\label{eq:ImPiREx}
S_1(q,\omega) =R_1(q,\omega) \text{I}(q,\omega) \cos^2[\delta(q,\omega,B)],
\end{eqnarray}
where
\begin{align}\label{eq:I}
\text{I}(q,\omega)&=\frac{ 2\big[(\omega^2-q^2)^2 +  4B^2 q^2\big]}{(\omega^2-q^2)^2  \sqrt{1-4B^2/(\omega^2-q^2)}}, \\
\delta(q,\omega,B)&=\frac{1}{2}\Big[\tan^{-1}\Big(\frac{p_{+}}{2B}   \Big)  +   \tan^{-1}\Big(\frac{p_{-}}{2B}   \Big)   \Big]\label{eq:delta},
\end{align}
with $p_{\pm}=|q|  \pm \omega\sqrt{1- 4B^2 /(\omega^2 -q^2)}$. The kinematic thresholds are encoded in the function
\begin{align}
  R_1(q,\omega) = \Theta (\omega^{\mathrm{Intra}}_\mathrm{U}-\omega)
  \Theta (\omega-\text{max}[\omega^{\mathrm{Intra}}_\mathrm{L+},\omega^{\mathrm{Intra}}_\mathrm{L-} ]),
\end{align}
where $\omega^{\mathrm{Intra}}_\mathrm{U} =\epsilon_{+} (k_F+|q|)-\mu$ is the upper threshold for intra-band processes, and $\omega^{\mathrm{Intra}}_\mathrm{L\pm} =\pm[  \epsilon_{+} (k_F - | q|)-\mu ] $ are the lower thresholds for $|q| \gtrless 2k_F$, as shown in Fig.~\ref{fig:DSF}a.

\begin{figure}[t]
             \includegraphics[width=7cm]{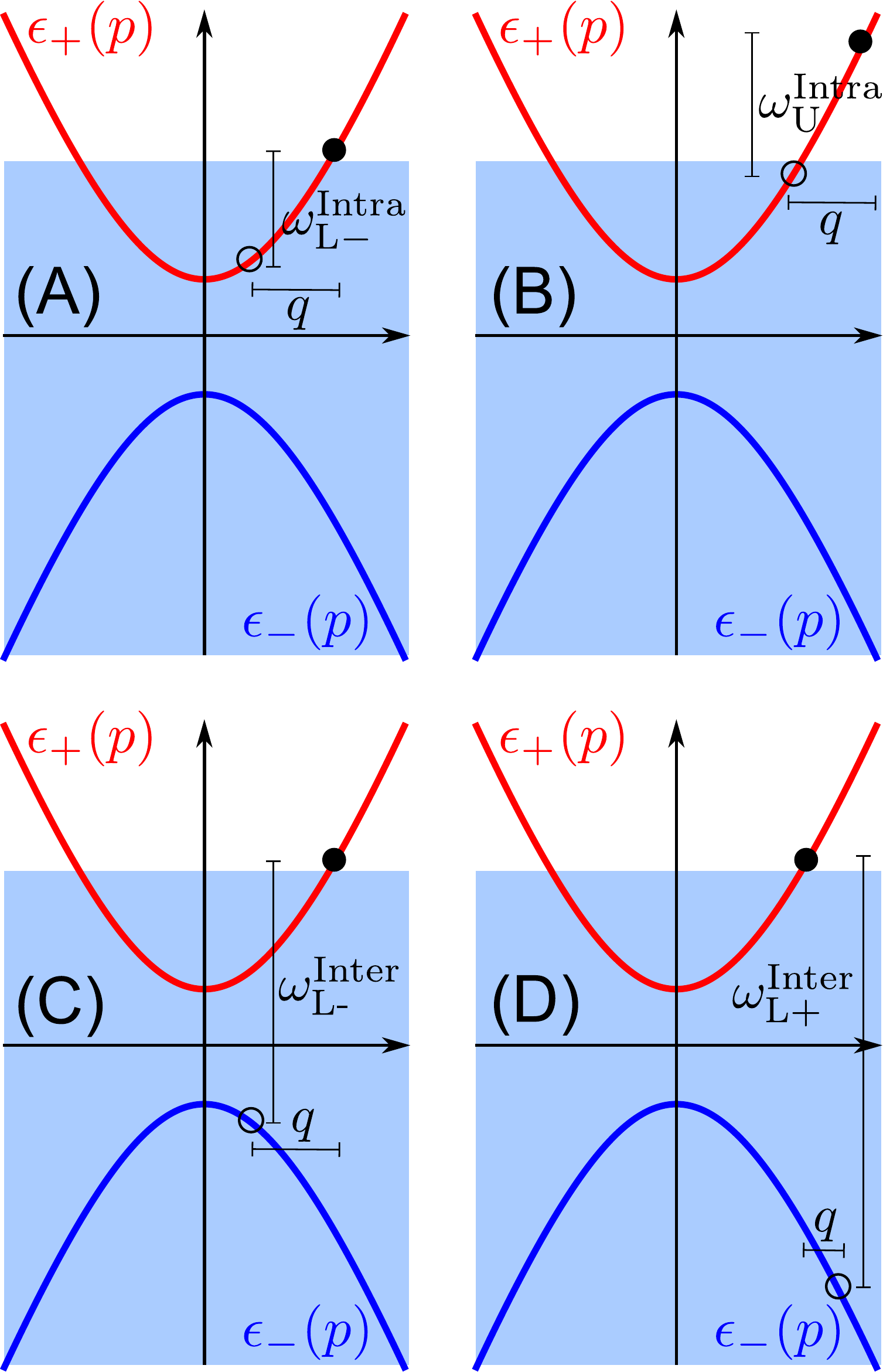}
    \centering
    \caption{The figures (A) to (D) illustrate the transitions giving rise to singularities at the thresholds $\omega^{\mathrm{Intra}}_\mathrm{L-},~\omega^{\mathrm{Intra}}_\mathrm{U},~\omega^{\mathrm{Inter}}_\mathrm{L+},~\text{and}~\omega^{\mathrm{Inter}}_\mathrm{L-}$ (see Fig.~\ref{fig:DSF}a for the edge thresholds). In the above figures the lower band $\epsilon_{-}(p)$ is completely filled, whereas the upper band, $\epsilon_{+}(p)$, is filled up to the chemical potential $\mu > B$ (indicated by the top of the blue shaded area).}
    \label{fig:DSFprocesses}
\end{figure}

Typical intra-band excitations involve transitions of a particle with momentum
$p_i$ and energy $\epsilon_+(p_i)$ from within the Fermi sea ($|p_i|<k_F=\sqrt{\mu^2-B^2}$) to a state with momentum $p_f = p_i +q$ and energy $\epsilon_+(p_f) = \epsilon_+(p_i) +\omega$ above the Fermi sea.  For $|q|<k_F$ the final state momentum $p_f$ will be parallel to the initial momentum ($p_i p_f > 0$;  see  for example Figs.~\ref{fig:DSFprocesses}A and~\ref{fig:DSFprocesses}B), whereas for $|q|>  k_F$, $p_f$ can be either parallel or anti-parallel to $p_i$ depending on whether $\omega  \gtrless  \sqrt{k_F^2 +B^2} -B $,
  respectively. Note that the anti-parallel option is absent for $B=0$, because the spin-states of the left and right moving fermions are then orthogonal.

It is known that for a quadratic spectrum $\epsilon =p^2/2m$, the noninteracting DSF at fixed $|q|< 2k_F$ is constant, $S(q,\omega) = m/|q|$, between the thresholds $\omega_\pm(q) = k_F| q|/m \pm q^2/(2m)$. In contrast, for the nonparabolic spectrum $\epsilon_\pm(p)$, the DSF exhibits power-law behavior, see Fig.~\ref{fig:DSF}b. For $B\ll \mu$ and $|q|< k_F$ it can be approximated as $S(q,\omega) \approx  4Bq^2/(q^2-\omega^2)^{3/2} $   with the width of the support  given by $\delta \omega= \omega^{\mathrm{Intra}}_\mathrm{U}-\omega^{\mathrm{Intra}}_\mathrm{L-} \approx  B^2q^2/\mu^3 $.

\emph{Inter-band transitions}: An inter-band process involves the excitation of a particle with momentum $p_i$ from the lower band to a state with momentum $p_f = p_i + q$ in the upper band. The energy difference is $\omega = \epsilon_+(p_f) - \epsilon_-(p_i)$. The corresponding contribution to the DSF is
\begin{align}\label{eq:Intra}
    S_2(q,\omega) = R_2(q,\omega) \text{I}(q,\omega) \sin^2[\delta(q,\omega,B)],
\end{align}
where
\begin{align}\label{ll-inter}
    R_2(q,\omega) &=\Theta (\omega -\omega^{\text{Inter}}_{L-}) +  \Theta (\omega -\omega^{\text{Inter}}_{L+}  )  \\
    & + 2 \Theta(q-2k_F) \Theta(\omega^{\text{Inter}}_{L-}-\omega)\Theta(\omega-\sqrt{q^2+ 4B^2}).\notag
\end{align}
Here the  threshold frequencies are at $\omega^{\text{Inter}}_{L\pm } = \mu +\epsilon(k_F \pm |q|)$ (see Fig.~\ref{fig:DSF} and  Figs.~\ref{fig:DSFprocesses}C,D for
the
threshold  transitions corresponding to $\omega^{\text{Inter}}_{L- } $ and $\omega^{\text{Inter}}_{L+ } $, respectively).

For  $\text{max}[\omega^{\text{Inter}}_{L-},B+ \sqrt{B^2+q^2}] < \omega < \omega^{\text{Inter}}_{L+}$  transitions with only  $p_i q > 0$ are allowed, whereas for frequencies $\omega > \omega^{\text{Inter}}_{L+} $ an additional channel involving anti-parallel momenta $p_i q < 0$ (however, $p_i p_f > 0$) opens up. Hence, the DSF acquires a discontinuity at $\omega =  \omega^{\text{Inter}}_{L+} $. For small momenta $|q|\ll k_F$ and energies $\omega \approx 2 \mu$, $S(q,\omega) \propto q^2 B^2/ (\omega^3 \sqrt{\omega^2- 4B^2})$.

On the other hand, transitions with anti-parallel initial and final momenta $p_i p_f < 0$ are possible for $|q|>k_F$.
For  $2k_F> |q|>k_F $ such transitions are allowed for  frequencies within
$\omega^{\text{Inter}}_{L-} < \omega  < B+ \sqrt{B^2+q^2}$.
However, for  $|q|> 2k_F$  the
 frequency  regime as   given in the second line of
 Eq.~(\ref{ll-inter}), i.e., $ \sqrt{q^2+4 B^2} < \omega  < \omega^{\text{Inter}}_{L-}$,   also
  supports  $p_i p_f < 0$ transitions.
We would like to point out that in this regime
the frequency $\omega=\sqrt{(p_i+q)^2 + B^2} +  \sqrt{(p_i)^2 + B^2} $ exhibits a nonmonotonic
behavior as a function of  $p_i$, with the minimum acquired at $p_i =-q/2$. The consequence of this is that
 $S(q,\omega)$ develops a square-root  singularity at the frequency $\omega =\sqrt{q^2+4 B^2}$.

\emph{Interactions:}
The edge features of the noninteracting structure factor are strongly modified even for weak interactions, since they generally lead to power-law singularities at the thresholds \cite{imambekov}. In the following we will discuss these in the long wavelength limit ($0 < q<  k_F$).

Interactions in 1D systems are typically treated via the Tomonaga-Luttinger liquid approach, which relies
on a linearization of the fermionic spectrum to express the degenerate  particle-hole excitations in terms of bosonic
density excitations. In contrast, as we saw already in the noninteracting case, the nonlinearity of the spectrum is essential for a proper description of the DSF at nonzero energies.
Unfortunately, it is not easy to incorporate the nonlinearity in the spectrum within this approach,
because the curvature terms lead to interactions between the bosonic modes. Perturbation theory in these interactions produces divergences and a proper resummation  is highly non-trivial~\cite{imambekov}.
Nevertheless, the  modifications of threshold features in the response function due to the interactions
fall under the paradigm of the Fermi edge singularity problem and are hence tractable~\cite{schotte,balents,pustilnik,perira,zvonarev,mishchenko,karimi,matthews,imambekov09,imambekov}. While spin-charge separation complicates the problem for conventional spinful fermions \cite{schmidt10,schmidt10a}, helical edge states involve the same number of degrees of freedom as spinless electrons. In that respect, the theory is more similar to that of spinless electrons \cite{imambekov09}. However, the facts that (i) the spectrum $\epsilon_\pm(p)$ is not parabolic and (ii) right-movers and left-movers have different spins lead to a result that is very different from that of spinless fermions.

Close to a given threshold energy, the transitions which determine the structure factor involve, e.g., a particle at a momentum $p$ ($|p| < k_F$) which is excited to one of the Fermi points. The interaction of the resulting ``deep hole'' with low-energy particles close to the Fermi points gives rise to singular features at the thresholds.
The approach to tackle this problem involves projecting the Hamiltonian onto three narrow bands~\cite{pustilnik}, two of which are
centered about the Fermi points, whereas the third one is centered about the deep hole momentum $p$.

We will first consider $S(q,\omega)$ at the threshold frequency $\omega \approx \omega^{\text{Intra}}_{\text{L}-}$ for $0 < q < k_F$, where the threshold configuration contains a deep hole at momentum $k_F - q$ as shown in Fig.~\ref{fig:DSFprocesses}A. In terms of the diagonal basis of the Hamiltonian (\ref{H1}),   the relevant  field operators  are
\begin{align}
\Psi(x) \approx\hat{r}   R(x) e^{ik_Fx} + \hat{l}  L(x)  e^{-ik_Fx}  +  \hat{u}(p) \mathcal{R}_{u}(x)     e^{ip x}   ,\label{eq:Field-RL}
\end{align}
where the operators $R$,  $L $ and $\mathcal{R}_u$ are slow degrees of freedom about the points $k_F$, $-k_F$
and $p$, respectively.   In terms of
$\hat{u}(p) =  \{ \cos(\gamma_p/2) ,-\sin(\gamma_p/2)\} ^{\text{T}}$, where  $\gamma_p=\tan^{-1}(B/p)$, the spinors represented by the
\emph{hat}  terms are $\hat{r}  = \hat{u}(k_F) $, $\hat{l}  = \hat{u}(-k_F) $.
Within each band, the non-interacting Hamiltonian can be linearized,
\begin{align}
H_0 = \int dx  \Big[ -i v R^\dagger \partial_x R + i v L^\dagger\partial_x L
-  \mathcal{R}_u^\dagger (\tilde{\omega} +i\tilde{v}\partial_x) \mathcal{R}_u \Big],
\end{align}
where $v=\partial  \epsilon_+/\partial k |_{k=k_F}$, $\tilde{v}=\partial  \epsilon_+/\partial k |_{k=p}$ is the velocity of the deep hole at momentum $p = k_F - q$, and $\tilde{\omega} =\omega^{\text{Intra}}_{\text{L}-}$.
Interactions are described by the usual density-density
interaction term,
$H_{\text{int}}=(1/2)\int dx dx' V(x-x') \rho(x)\rho(x')$, where  the density operator
is $\rho(x)= \Psi^\dagger(x)\Psi(x)$.
 In terms of the two bosonic fields $\phi$ and $\theta$, which satisfy the canonical commutation
relation $[\phi(x),\partial_y \theta(y)]=i \Theta(x-y)$,
  the right and left  moving fields near the fermi level acquire
the form $R(x) = e^{i\sqrt{\pi}(-\phi+\theta)}/\sqrt{2\pi a_0}$ and
 $L(x) = e^{i\sqrt{\pi}(\phi+\theta)}/\sqrt{2\pi a_0}$, respectively, where $a_0$ is a short-distance cutoff. We can now decompose the  full Hamiltonian $H= H_0 + H_{\text{int}}$ into three parts
\begin{align}\label{eq:FullH}
H=\int dx \mathcal{R}_u^\dagger(-\tilde{\omega} -i\bar{v}\partial_x) \mathcal{R}_u+H_{\text{LL}} + H_{M},
\end{align}
where $\bar{v} =  \tilde{v} + V(0) /2\pi $ is the renormalized impurity velocity, and $V(q)$ denotes the Fourier transport of the interaction potential $V(x)$. The second term is a conventional Luttinger liquid Hamiltonian, $ H_{\text{LL}}=  (1/2) \int dx [uK(\partial_x\theta)^2   + (u/K) (\partial_x\phi)^2] $, where $uK=v[1+   V(2k_F)\sin^2\gamma_{k_F}/(2\pi v)  ]$ and $u/K=v[1+ V(0)/(\pi v)-V(2k_F)  \sin^2\gamma_{k_F}/(2\pi v)  ]$. Here, $u$ denotes the sound velocity of the Luttinger liquid and $K$ is the Luttinger parameter. The third term in Eq.~(\ref{eq:FullH})  mixes the deep hole with the bosonic modes
 \begin{align}\label{eq:Int4}
H_M&=\sqrt{\frac{1}{4\pi}} \int dx   \mathcal{R}_u^\dagger\mathcal{R}_u \Big(\alpha_1 \partial_x \phi  + \beta_1  \partial_x \theta\Big),
\end{align}
where, using $\gamma_\pm = (\gamma_{k_F-q} \pm\gamma_{k_F}$)/2,
\begin{align}\label{eq:Coefficients}
\alpha_1 &= 2V(0) -V(q) \cos^2\gamma_-  -V(2k_F-q)\sin^2\gamma_+{}\nonumber\\
\beta_1 &=  V(q) \cos^2\gamma_- -V(2k_F-q)\sin^2\gamma_+.
\end{align}

Determining $S(q,\omega)$ for $\omega \approx \omega^{\text{Intra}}_{\text{L}-}$
requires a calculation of the correlation function
$ \langle  \mathcal{R}_u^\dagger (x) R(x) \mathcal{R}_u(0)R^\dagger(0)\rangle$ in Fourier space.
 To this end, we perform a unitary transformation on the Hamiltonian~(\ref{eq:FullH}) in order to remove the terms linear in $\partial_x \phi$
 and $\partial_x \theta$~\cite{pustilnik,imambekov}. The  transformation also modifies the fields in the correlation function, but evaluating
 it is straightforward  and we obtain
\begin{align}\label{eq:Int}
S(q,\omega)\propto  \int  \frac{dt e^{i(\omega-  \tilde{\omega})  t} }{(\epsilon + iu t - i \bar{v}t)^{\delta_R^2/4\pi}    (\epsilon + iut + i \bar{v}t)^{\delta_L^2/4\pi}   },
\end{align}
where $\epsilon = 0^+$,  $\delta_L= \sqrt{\pi K} - \sqrt{\pi/K}$ and $\delta_R= \sqrt{\pi K} + \sqrt{ \pi/K} + (\alpha_1\sqrt{K}-\beta_1/\sqrt{K})/[\sqrt{4\pi}(\bar{v}-u)]$. Since  $u>\bar{v}$ the integral is non-zero only for $\omega>\tilde{\omega}$. Therefore, we obtain for $\omega \approx \tilde{\omega}\equiv\omega_{\text{L}-}^{\text{Intra}}$,
\begin{align}\label{eq:struct-Lower}
S(q,\omega) \propto \frac{\theta[\omega -\tilde{\omega}(q)]}{[\omega -\tilde{\omega}(q)]^\nu},
\end{align}
The exponent $\nu = 1- (\delta_R^2 + \delta_L^2)/4\pi$ becomes for weak interactions,
\begin{align}\label{eq:expo}
\nu  =\frac{V(0) -V(q)\cos^2\left[(\gamma_{k_F-q}-\gamma_{k_F})/2\right]}{  \pi (u-\bar{v})} > 0.
\end{align}
where $p = k_F - q$. Therefore the structure factor diverges for frequencies above $\omega^{\text{Intra}}_{\text{L}-}$, and vanishes below this threshold.

A similar analysis yields the behavior near the  threshold $\omega_{\text{L}+}^{\text{Intra}}$. Now the
 velocity $\bar{v}$ at $p=k_F+q$ is greater than  $u$. This has two main consequences. On the one hand, the exponent $\nu$, which is still formally given by Eq.~(\ref{eq:expo}), now has its sign reversed, $\nu <0$, so the DSF is convergent at this threshold. On the other hand, the integral (\ref{eq:Int}) corresponding to
 $\tilde{\omega}=\omega_{\text{L}+}^{\text{Intra}}$ is non-zero on either side of  $\omega_{\text{L}+}^{\text{Intra}}$.

The DSF near the lower threshold for inter-branch processes, $\omega_{\text{L}-}^{\text{Inter}}$, exhibits a one-sided divergence,
$S(q,\omega) \propto \theta[\omega -\omega_{\text{L}-}^{\text{Inter}}(q)]/[\omega- \omega_{\text{L}-}^{\text{Inter}}(q)]^{\nu'}$.
 The   exponent is  now modified to
$\nu' = (V(0) -V(q)\sin^2[(\gamma_{k_F-q}-\gamma_{k_F})/2])/[ \pi (u-\bar{v}) ]$,   where     $\bar{v}<0$
 is the velocity of the fermion in the lower band.  Near  the
threshold $ \omega_{\text{L}+}^{\text{Inter}}$  (due to transition of a fermion with momentum $k_F+q$ from the lower band to the Fermi level  $k_F$ in the upper band)
the exponent still has the above  $\nu'$ form, but the response function exhibits divergences from both  sides of the threshold frequency
 since in this regime  $|\bar{v}| > u$.

 \begin{figure}[t]
        \includegraphics[width=\columnwidth]{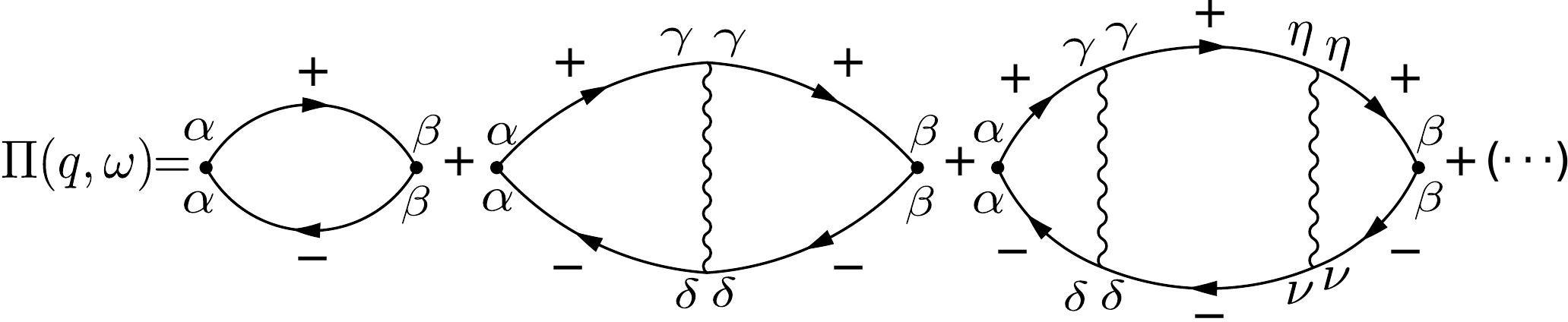}
    \centering
    \caption{Ladder diagram resummation for $|\mu| < B$. Solid lines with $\pm$ denote single-particle Green's function corresponding to the upper and lower bands, respectively.  The greek letters represent spin indices (repeated indices imply summation), and  wiggly lines denote interactions. Note the first term in the series is given by Eq.~(\ref{PO1}).}
   \label{fig:ladder}
\end{figure}

So far we have considered the case $\mu > B$. Due to the particle-hole symmetry, the structure factor exhibits identical behavior for $\mu < -B$.
On the other hand, a  different scenario emerges for $|\mu|< B$. The chemical potential   now lies in the gap  and  the non-interacting  response function
exhibits square-root singularity  at the edge  $\omega= \sqrt{q^2 + 4B^2 }$.  As the chemical potential is in the gap, interactions can no longer be treated  via the Luttinger liquid theory. However, exploiting the similarity to the problem of Mahan excitons in semiconductors \cite{mahan}, we used a ladder diagram resummation (see Fig.~\ref{fig:ladder}) for a generic interaction potential $V(q)$ to map the problem on a single-particle Schr\"odiner equation. As a result, one finds sharp (delta-function type) subgap resonances in the DSF due to the formation of two-particle bound states.

To simplify the result, we will assume a contact potential between electrons, $V(q) = V(0)$, in the following. In this case the polarization operator can be
evaluated by directly summing up the ladder series in Fig.~\ref{fig:ladder}.
We obtain
\begin{align}\label{Pi-T2}
S(q, \omega)&=\frac{\pi q^2}{2B^{3/2}}  \frac{\sqrt{\omega - \omega_q}  }{ \omega - \omega_q   +\pi^2 V(0)^2 B}\Theta (\omega -  \omega_q) {}\nonumber\\
 &+\frac{\pi^3 q^2 V(0)}{B}\delta[\omega -\omega_q +  \pi^2 V(0)^2B],
  \end{align}
where $\omega_q = 2B +q^2 /4B$ (note for  $|q|\ll B$, $\sqrt{4B^2+q^2} \approx \omega_q  $). From Eq.~(\ref{Pi-T2}), we can draw two important conclusions. First, the interactions modify the
square root divergence at $\omega= \omega_q$  (present in the non-interacting limit) into a square-root suppression.
The second nontrivial effect  is the emergence of a single bound-state resonance, which manifests itself as a sharp peak in the structure factor at sub-gap energies, $\omega = \omega_q -  \pi^2 V(0)^2B$.

The structure factor and its accompanying singular features can in principle be extracted based on  the  recently proposed technique
involving a source-probe setup~\cite{stano}. The proposal would be to use  time-dependent electric  field at the source point to create charge
excitations in the 1D helical modes which would then  induce currents at the probe point, thus yielding information
on the spatially and temporally  resolved  response function.   In addition, Coulomb drag measurements can  serve as a useful probe for the DSF~\cite{pustilnik03,yamamoto}.

To summarize, we have studied the dynamical structure factor $S(q,\omega)$ of a helical liquid, and the role of magnetic-field induced nonlinear spectrum. We explicitly
 considered  the  contributions from the intra- and inter-band   transitions. We found that the thresholds present in the noninteracting  $S(q,\omega)$ turn into power-law singularities upon the introduction of interactions. The edge exponents $\nu$
 depend on the momentum $q$, the interaction strength, and the curvature of the spectrum. As a consequence of the nonparabolic spectrum and the nontrivial spin texture of the edge states, the DSF differs strongly from that of conventional spinless and spinful fermion systems.

\acknowledgments
S.G. gratefully acknowledges   the hospitality of the Department of Physics at the University of Basel.
The work of T.L.S. and D.L. was supported by the Swiss NSF, NCCR Nanoscience, and NCCR QSIT.

\end{document}